\documentclass[10pt, twocolumn]{IEEEtran}

\usepackage{graphicx}
\usepackage{color}

\usepackage{amsmath}
\usepackage{amssymb}
\usepackage{amsthm}
\usepackage{arydshln}    
\usepackage{bm}
\usepackage{cite}

\usepackage[caption=false,font=footnotesize]{subfig}

\newtheorem{theorem}{Theorem}
\newtheorem{corollary}[theorem]{Corollary}

\def\MID{\, | \,}
\def\P{\mathbb{P}}
\def\Hhat{\widehat{H}}
\def\Belief{q}

\newcommand{\LocalI} [1]{{P_{e,#1}^{\rm I}}}
\newcommand{\LocalII}[1]{{P_{e,#1}^{\rm II}}}
\newcommand{\LocalIx}{{P_{e}^{\rm I}}}
\newcommand{\LocalIIx}{{P_{e}^{\rm II}}}
\newcommand{\LocalIxSeq}[1]{{P_{e}^{{\rm I}_{#1}}}}
\newcommand{\LocalIIxSeq}[1]{{P_{e}^{{\rm II}_{#1}}}}
\newcommand{\LocalISeq}[3]{{P_{e,#1_{#3}}^{{\rm I}_{#2}}}}    
\newcommand{\LocalIISeq}[3]{{P_{e,#1_{#3}}^{{\rm II}_{#2}}}}
\newcommand{\GlobalI} {{P_{E}^{\rm I}}}
\newcommand{\GlobalII}{{P_{E}^{\rm II}}}

\begin{document}

\title{Keep Ballots Secret: \\ On the Futility of Social Learning in
         Decision Making by Voting}
\author{Joong Bum Rhim and Vivek K Goyal%
\thanks{This material is based upon work supported by the National Science Foundation under Grant No.~1101147.}
\thanks{The authors are with the Research Laboratory of Electronics,
  Massachusetts Institute of Technology, Cambridge, MA 02139 USA.}
}

\maketitle

\begin{abstract}
We show that social learning is not useful in a model of team
binary decision making by voting, where each vote carries equal weight.
Specifically, we consider Bayesian binary hypothesis testing
where agents have any conditionally-independent
observation distribution and their local decisions are fused by
any $L$-out-of-$N$ fusion rule.
The agents make local decisions sequentially, with each allowed to
use its own private signal and all precedent local decisions.
Though social learning generally occurs in that precedent local decisions
affect an agent's belief, optimal team performance is obtained when
all precedent local decisions are ignored.
Thus, social learning is futile, and secret ballots are optimal.
This contrasts with typical studies of social learning because we include
a fusion center rather than concentrating on the performance of the
latest-acting agents.
\end{abstract}


\section{Introduction}    \label{sec:Introduction}

Consider a set of agents making a decision collectively by voting,
with each agent having equal influence.
It would seem that the best collective decisions would come from each
agent having as much information as possible.
For example, if the voting is sequential rather than simultaneous,
it would seem that the collective decision would be improved by
later-acting agents having knowledge of
the votes of the earlier-acting agents.
Using the methods of signal processing to analyze a simple model,
we show that this is not the case.
In a social or political context,
this provides a mathematical
justification for the use of secret ballots,
independent of any other merits, such as avoiding bribery, intimidation,
or insincere voting.
To be clear, though, our interest is in binary decision making under uncertainty,
not voting to express individual preferences.

In contemporary popular culture,
the advantage of (properly) aggregating opinions of many
(reasonably diverse and independent) agents is known as
``the wisdom of crowds''~\cite{Surowiecki2004}.
In statistical signal processing, this is nothing more than
a reduction in effective noise level from averaging or aggregating.
Suppose \emph{private signals} observed by decision-making agents are not perfectly informative due to noise.
If $N$ agents share their private signals, assumed conditionally independent and identically distributed (iid) given the unknown variable of interest, then the sharing effectively increases the signal-to-noise ratio (SNR) by a factor of $N$.  Even if a communication capacity limit restricts the amount of shared information, sharing still helps the detection system work much better than a single-agent system.
For example, in systems where agents with Gaussian noise-corrupted
observations share 1-bit signals
that are fused with the {\sc majority} fusion rule, the effective SNR is asymptotically increased
 by a factor of $2N / \pi$~\cite{VarshneyRVG2011}.

In this paper, we consider a team\footnote{The term \emph{team} implies that the agents have no
conflicts of interest, which here translates to agreement in the relative
importance of false alarms and missed detections~\cite{Radner1962}.} of agents that
together make a binary decision (between 0 and 1) by voting.
Binary decisions often made by teams include
whether to make a large purchase,
whether to hire a job applicant,
and whether to convict a defendant;
the final example highlights that aggregation by voting,
with equal weight to each vote,
does not imply the use of majority rule.
The agents have their own private signals and
make local decisions in some preordained order.
Because of this ordering, we refer to the agents as a sequence.
The agents form a distributed detection and data fusion system because their local decisions are fused to reach a global decision, and only the global decision determines their performance.

Now assume that the agents can watch other agents' choices as \emph{public signals} so that they can use the predecessors' decisions in decision making.
The framework of sequential decision making was independently introduced in~\cite{Banerjee1992} and~\cite{BikhchandaniHW1992}.
A key concept in these works is \emph{herding}.
A herd of agents takes place when all agents beyond some index
have the same local decision.
Both works found that a herd of agents making the incorrect decision
might arise with positive probability
when private signals are bounded; this is a disappointing
occurrence since the optimal aggregation of private signals would result
in vanishing error probability.  Subsequently, \cite{SmithS2000} showed that agents will asymptotically settle on the correct decision if private signals are unbounded.  Recently, \cite{AcemogluDLO2011} extends the result to general network topologies where each agent can observe decisions made by its neighbors instead of all previous agents.

This work contrasts from previous works in two key ways:
we consider an arbitrary but finite number of agents $N$; and
we consider any symmetric fusion of the local decisions (i.e., voting
by an $L$-out-of-$N$ rule\footnote{In the notation of $L$-out-of-$N$ rule, $N$ denotes the number of agents and $L$ denotes the minimum number of
1 votes for 1 to be the global decision.}).
Because the number of agents is finite, presence or absence of herding
is not central to our study; a herd of agents might arise and yet form
a small fraction of the $N$ total agents.
Also, the fusion by voting makes an early-acting agent important not
only for its influence on later-acting agents but also because its
vote counts.

As the first to study this scenario, we ask whether
it is beneficial for the agents not only to send their local decisions to the fusion center but also to share among themselves.

The public signals raise two changes to the model:
the belief update that is standard in social learning
and a fusion rule update that has not appeared previously
in the context of social learning.

Each agent will update its belief about the true state based on the public signals (in a Bayesian setting) and design a better decision rule.  The previous decisions reflecting the private signals of the precedent agents contain information that the following agents do not have.  If an agent observes that most predecessors have chosen 1, then the agent will have a stronger belief on 1 and a weaker belief on 0\@.  Hence, the update of belief is a positive feedback that encourages an agent to follow the precedents.

Each agent will adjust the fusion rule based on the public signals.  For example, if the first agent has chosen 0, then the $L$-out-of-$N$ fusion rule changes to the $L$-out-of-$(N{-}1)$ rule for the remaining agents.  And if the second agent has subsequently chosen 1, then the fusion rule becomes the $(L{-}1)$-out-of-$(N{-}2)$ rule, etc.  Thus, the previous decisions put a following agent in a different position; if most previous agents have chosen 1, then a few more votes for 1 will determine the global decision as 1\@.  This fact makes the agent become more careful to choose 1\@.  The evolution of the fusion rule is a negative feedback that discourages an agent to follow the precedents.

Our mathematical analysis shows that those positive and negative effects are exactly canceled out and that the optimal decision rules with public signals are equal to those without public signals.  It implies that observing public signals is practically useless.

Section~\ref{sec:Model} describes our decision-making model
formally.  Section~\ref{sec:2-agent} provides results for two agents and Section~\ref{sec:N-agent} extends the results for an arbitrary number of agents.  Section~\ref{sec:Conclusion} concludes the paper.

\section{The Formal Model}    \label{sec:Model}

A team of $N$ agents---Alexis, Blake, ..., Norah---performs a binary hypothesis test for an object.  The agents are aware of the prior probabilities of the object in state $H = 0$ and in state $H = 1$, which are denoted by $p_0 = \P\{H = 0\}$ and $p_1 = \P\{H = 1\} = 1 - p_0$.  The agents individually receive private signals about the state and perform tests to make their own binary decisions $\Hhat_i$.  They are a team in the sense that they place the same relative importance on
false alarms and missed detections and
share a common goal to minimize the cost of global decisions.  The cost for a false alarm is $c_{10}$ and the cost for a missed detection is $c_{01}$.  Thus, for the $L$-out-of-$N$ fusion rule, the average cost (i.e., Bayes risk) is given as follows:
\begin{align}
    R & = c_{10} p_0 \P\{\textstyle \sum_{i = 1}^{N} \Hhat_i \geq L \MID H = 0\} \nonumber \\
    & \quad + c_{01} p_1 \P\{\textstyle \sum_{i = 1}^{N} \Hhat_i \leq L - 1 \MID H = 1\}.
\end{align}

The private signals $Y_i$ are conditionally iid given the state $H$, with the likelihood function $f_{Y_i | H}$.  In other words, all agents observe private signals of equal quality.  Bayesian agents use a likelihood ratio test (LRT) as the optimal decision rule, which chooses $\Hhat_i = 1$ if the ratio $f_{Y_i | H}(y_i \MID 1) / f_{Y_i | H}(y_i \MID 0)$ is greater than a certain threshold~\cite{NeymanPearson1933,Varshney97}.  We assume that the likelihood ratio is an increasing function of $y_i$ so that the LRT is simplified to a decision rule with a decision threshold $\lambda_i$:
\begin{equation}
    y_i \overset{\Hhat_i(y_i) = 1}{\underset{\Hhat_i(y_i) = 0}{\gtreqless}} \lambda_i.
\end{equation}

We will compare the optimal decision thresholds in the following two scenarios:
\begin{itemize}
    \item[1)] A common distributed detection system:  Agents only observe private signals and make local decisions in parallel without knowing other agents' decisions.  Then their decisions are fused by a specific $L$-out-of-$N$ rule to make a global decision.  Alternatively, we call this scenario parallel decision making.
    \item[2)] A distributed detection system combined with sequential decision making:  Agents observe private signals and sequentially make local decisions.  Before each agent makes a decision, the agent observes precedents.  Then their decisions are fused by a specific $L$-out-of-$N$ rule to make a global decision.
\end{itemize}
Our notations for decision thresholds are distinguished in the two scenarios:  $\lambda$ in the first scenario and $\rho$ in the second scenario.

In Scenario 1, we will say that the optimal decision thresholds are identical, i.e., $\lambda_1^{\ast} = \cdots = \lambda_N^{\ast}$ for the following reasons:
Using identical decision thresholds is asymptotically optimum for the binary hypothesis testing problem~\cite{Tsitsiklis1988}.  Furthermore, by numerical experiments, it turns out that constraining to identical decision rules causes little or no loss of performance for finite $N$ and the corresponding optimal fusion rule has the $L$-out-of-$N$ form~\cite{Tsitsiklis1993}.  Our numerical experiments
(not reported here)
for fixed fusion rules and conditionally iid private signals show that the optimal decision thresholds are in fact identical at least for any $N \leq 7$ in multiple cases with such as Gaussian and exponential likelihood functions.

We use $\LocalIx$ and $\LocalIIx$ as notations for probabilities of Type I (false alarm) and Type II (missed detection) errors.  We use a subscript index to indicate a specific agent.  For example, error probabilities of Alexis
are written as
\begin{equation}
    \LocalI{1} = \P\{\Hhat_1 = 1 \MID H = 0\}, \ \  \LocalII{1} = \P\{\Hhat_1 = 0 \MID H = 1\}.
    \label{eq:ErrorProb_NoPublicSignal}
\end{equation}
Since all agents have equally good private signals, their error probabilities are the same as \eqref{eq:ErrorProb_NoPublicSignal} in the parallel decision-making scenario.

In Scenario 2, the agents sequentially make hard decisions: Alexis first makes a decision $\Hhat_1$, Blake next makes a decision $\Hhat_2$, and so on.  Predecessors' decisions, which we call public signals, may cause different error probabilities by following agents.  In this case, we use superscript notation to specify the public signals.  For example, $\LocalISeq{2}{0}{}$ denotes the probability of Type I error of Blake (the second agent) upon observing $\Hhat_1 = 0$ and $\LocalIISeq{3}{01}{}$ denotes the probability of Type II error of Chuck (the third agent) upon observing $\Hhat_1 = 0$ and $\Hhat_2 = 1$.  We sometimes use a notation like $\LocalISeq{3}{0}{}$ that specifies Alexis's decision as $\Hhat_1 = 0$ but does not specify Blake's decision $\Hhat_2$.

Once the error probabilities of all agents are computed, the conditional probabilities of the global decision being in error are given from the $L$-out-of-$N$ fusion rule:
\begin{align}
    \GlobalI & = p_{\Hhat | H}(1 \MID 0) = \sum_{n = L}^{N} \sum_{I \subseteq [N] \atop |I| = n} \prod_{i \in I} \LocalI{i} \prod_{j \in [N] \setminus I} \left(1 - \LocalI{j}\right),\nonumber \\
   \GlobalII & = p_{\Hhat | H}(0 \MID 1) = \sum_{n = N{-}L{+}1}^{N} \sum_{I \subseteq [N] \atop |I| = n} \prod_{i \in I} \LocalII{i} \prod_{j \in [N] \setminus I} \left(1 - \LocalII{j}\right), \nonumber
\end{align}
where $[N]$ denotes the set $\{1, 2, \ldots, N\}$.
The Bayes risk is based on the team decision:
\begin{align}
    R & = c_{10} p_0 \GlobalI + c_{01} p_1 \GlobalII.
    \label{eq:BayesRisk}
\end{align}

\section{Two Agents}    \label{sec:2-agent}

\begin{figure}
    \centering{
    \subfloat[]{\includegraphics[width=1.4in]{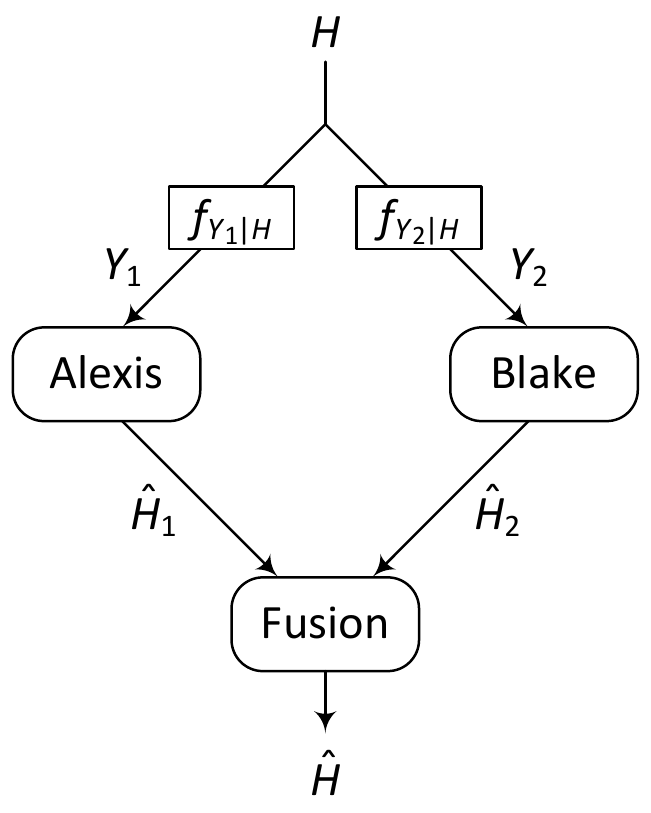}}
    \hfil
    \subfloat[]{\includegraphics[width=1.4in]{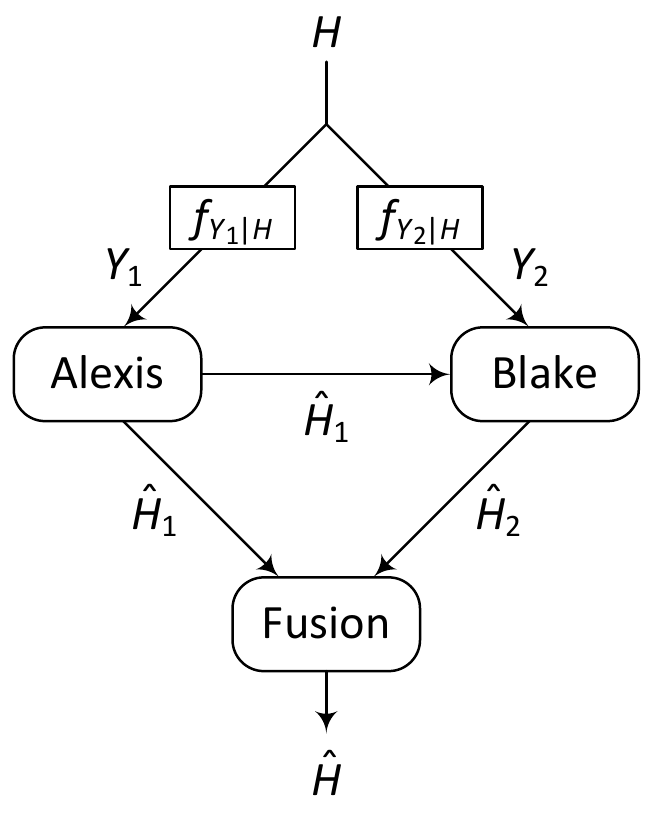}}
    }
    \caption{Two agents and a fusion center. (a) Parallel decision making.  (b) Sequential decision making.}
    \label{fig:TwoAgents}
\end{figure}

Let us consider the simplest case for distributed detection, which is $N = 2$.  We will compare two cases, which are depicted in Fig.~\ref{fig:TwoAgents}.  We start from an assumption that Alexis does not change her strategy whether the agents observe public signals or not.  The assumption comes from an intuition that she is the first agent to make a decision and does not observe any previous decisions in both scenarios.

\begin{theorem}
    Suppose that sharing previous decisions does not change Alexis's decision rule.  Then, for $N = 2$ and any $L$-out-of-$N$ fusion rule, the existence of public signals does not affect the decision strategies of agents.
    \label{thm:2Agent}
\end{theorem}
\begin{IEEEproof}
    There are only two meaningful fusion rules when $N = 2$: 1-out-of-2 ({\sc or}) rule and 2-out-of-2 ({\sc and}) rule.  Let's consider the {\sc or} rule first.

    In the parallel decision-making scenario, the Bayes risk is given by
    \begin{align}
        R_p
        & = c_{10} p_0 \left( \LocalI{1} + \LocalI{2} - \LocalI{1} \LocalI{2} \right) + c_{01} p_1 \LocalII{1} \LocalII{2},
        \label{eq:BR,1-out-of-2,Parallel}
    \end{align}
    where Alexis's error probabilities $\LocalI{1}$ and $\LocalII{1}$ are governed by her decision threshold $\lambda_1$ and Blake's $\LocalI{2}$ and $\LocalII{2}$ are governed by his decision threshold $\lambda_2$.  Their optimal decision thresholds $\lambda_1^{\ast}$ and $\lambda_2^{\ast}$ are the minimizer of \eqref{eq:BR,1-out-of-2,Parallel}.

    In the sequential decision-making scenario, Blake observes Alexis's decision $\Hhat_1$.  He makes a decision carefully only if her decision is $\Hhat_1 = 0$ because $\Hhat_1 = 1$ already determines the global decision as $\Hhat = 1$.
Thus, we take care of Alexis's decision threshold, $\rho_1$, and Blake's decision threshold upon observing $\Hhat_1 = 0$, which is denoted by $\rho_2^{_{0}}$, only.  The Bayes risk is given by
    \begin{align}
        R_s
        & = c_{10} p_0 \left( \LocalISeq{1}{}{} + \left(1 - \LocalISeq{1}{}{}\right) \LocalISeq{2}{0}{} \right) + c_{01} p_1 \LocalIISeq{1}{}{} \LocalIISeq{2}{0}{},
        \label{eq:BR,1-out-of-2,Serial}
    \end{align}
    which is exactly the same formula as \eqref{eq:BR,1-out-of-2,Parallel}.  Therefore, the optimal decision thresholds would be the same as those in the first case:
    \begin{equation}
        \rho_1^{\ast} = \lambda_1^{\ast}, \qquad \rho_2^{_{0} \ast} = \lambda_2^{\ast},
    \end{equation}
    and $\rho_1^{\ast} = \rho_2^{_{0} \ast}$ because $\lambda_1^{\ast} = \lambda_2^{\ast}$.  Since we do not care about Blake's decision threshold when $\Hhat_1 = 1$, $\rho_2^{_{1}}$, we can set $\rho_2^{_{1} \ast} = \rho_2^{_{0} \ast}$, too.  Therefore, Alexis and Blake's optimal decision thresholds are not affected by the existence and the value of the public signal $\Hhat_1$.

    The same statement for the {\sc and} fusion rule can also be proven in a similar way.
\end{IEEEproof}

\section{$N$ Agents}    \label{sec:N-agent}

In Section~\ref{sec:2-agent}, we have shown that knowing Alexis's decision practically does not change Blake's optimal strategies for $N = 2$ and any $L$-out-of-$N$ fusion rule.  Now let us expand the problem to a general $N$-agent problem by mathematical induction.

\begin{theorem}
    Suppose that sharing previous decisions does not change Alexis's decision rule (i.e., $\rho_1^{\ast} = \lambda_1^{\ast}$).  If the existence of the public signals does not affect optimal decision thresholds of a team of $N$ agents for a specific $N$ and any $K$-out-of-$N$ fusion rule, then the existence of the public signals also does not affect optimal decision thresholds of a team of $N{+}1$ agents and any $L$-out-of-$(N{+}1)$ fusion rule.
    \label{thm:NtoN+1}
\end{theorem}

\begin{IEEEproof}
    First, consider the parallel decision-making scenario with $N{+}1$ agents.  Since a decision of an agent is critical only if the other $N$ local decisions are $L{-}1$ ones and $N{-}L{+}1$ zeros, the optimal decision threshold $\lambda^*$ is the solution to\footnote{Please see~\cite{RhimVG2012c} for a detailed description of how \eqref{eq:DecThres,N+1agents,parallel} is derived.}
    \begin{align}
        \frac{f_{Y | H} (\lambda \MID 1)}{f_{Y | H} (\lambda \MID 0)}
        & = \frac{c_{10} p_0 \binom{N}{L-1} \left(\LocalIx \right)^{L-1} \left(1 - \LocalIx \right)^{N-L+1}}{c_{01} p_1 \binom{N}{N-L+1} \left(\LocalIIx \right)^{N-L+1} \left(1 - \LocalIIx \right)^{L-1}} \nonumber \\
        & = \frac{c_{10} p_0 \left(\LocalIx \right)^{L-1} \left(1 - \LocalIx \right)^{N-L+1}}{c_{01} p_1 \left(\LocalIIx \right)^{N-L+1} \left(1 - \LocalIIx \right)^{L-1}},
        \label{eq:DecThres,N+1agents,parallel}
    \end{align}
    where we embed the idea that the optimal decision thresholds of all agents are identical.

    Next, consider the sequential decision-making scenario when the agents observe previous decisions.  The Bayes risk is given by
    \begin{align}
        R_p
        & = c_{10} p_0 \left( 1 - \LocalI{1} \right) \P\left\{\textstyle \sum_{n = 2}^{N+1} \Hhat_n \geq L \MID \Hhat_1 = H = 0\right\} \nonumber \\
        & \quad + c_{10} p_0 \LocalI{1} \P\left\{\textstyle \sum_{n = 2}^{N+1} \Hhat_n \geq L - 1 \MID \Hhat_1 = 1, H = 0\right\} \nonumber \\
        & \quad + c_{01} p_1 \LocalII{1} \P\left\{\textstyle \sum_{n = 2}^{N+1} \Hhat_n \leq L - 1 \MID \Hhat_1 = 0, H = 1\right\} \nonumber \\
        & \quad + c_{01} p_1 \left( 1 - \LocalII{1} \right) \P\left\{ \sum_{n = 2}^{N+1} \Hhat_n \leq L - 2 \MID \Hhat_1 = H = 1\right\} \nonumber \\
        & \triangleq R_0 \left(p_0 (1 {-} \LocalISeq{1}{}{} ) {+} p_1 \LocalIISeq{1}{}{} \right) + R_1 \left(p_0 \LocalISeq{1}{}{} {+} p_1 ( 1 {-}  \LocalIISeq{1}{}{} ) \right),
        \label{eq:BR,L-out-of-N+1,serial}
    \end{align}
    where $R_0$ and $R_1$ are specified in \eqref{eq:BR-B0} and \eqref{eq:BR-B1} and we define
    \begin{align}
        \Belief^{_0} & \triangleq \frac{p_0 \left(1 - \LocalISeq{1}{}{} \right)}{p_0 \left(1 - \LocalISeq{1}{}{} \right) + p_1 \LocalIISeq{1}{}{}} = \P\{H = 0 \MID \Hhat_1 = 0\}, \nonumber \\
        \Belief^{_1} & \triangleq \frac{p_0 \LocalISeq{1}{}{}}{p_0 \LocalISeq{1}{}{} + p_1 \left( 1 -  \LocalIISeq{1}{}{} \right)} = \P\{H = 0 \MID \Hhat_1 = 1\}.
        \label{eq:BeliefUpdate_N+1}
    \end{align}

    When the agents $2, \ldots, N{+}1$ observe that $\Hhat_1 = 0$, their optimal decision strategy is to minimize the term $R_0$ from \eqref{eq:BR,L-out-of-N+1,serial}:
    \begin{align}
        R_0 & = c_{10} \Belief^{_0} \P\left\{\textstyle \sum_{n = 2}^{N+1} \Hhat_n^{_0} \geq L \MID H = 0\right\}  \nonumber \\
        & \quad + c_{01}(1 - \Belief^{_0}) \P\left\{\textstyle \sum_{n = 2}^{N+1} \Hhat_n^{_0} \leq L - 1 \MID H = 1\right\},
        \label{eq:BR-B0}
    \end{align}
    where the condition $\Hhat_1 = 0$ is intended in the term $\Hhat_n^{_0}$.  Please note that $R_0$ is the same as the Bayes risk of $N$ agents when the prior probability is $\Belief^{_0}$ and fusion is by the $L$-out-of-$N$ rule.  It implies that the optimal decision thresholds of Agents $2, \ldots, N{+}1$ are the same as those of $N$ agents with prior probability $\Belief^{_0}$ and the $L$-out-of-$N$ fusion rule.

    Likewise, when the agents $2, \ldots, N{+}1$ observe that $\Hhat_1 = 1$, their optimal decision strategy is to minimize the term $R_1$ from \eqref{eq:BR,L-out-of-N+1,serial}:
    \begin{align}
        R_1 & = c_{10} \Belief^{_1} \P\left\{\textstyle \sum_{n = 2}^{N+1} \Hhat_n^{_1} \geq L - 1 \MID H = 0\right\}  \nonumber \\
        & \quad + c_{01}(1 - \Belief^{_1}) \P\left\{\textstyle \sum_{n = 2}^{N+1} \Hhat_n^{_1} \leq L - 2 \MID H = 1\right\}.
        \label{eq:BR-B1}
    \end{align}
    Their optimal decision thresholds are the same as those of $N$ agents with prior probability $\Belief^{_1}$ and $(L{-}1)$-out-of-$N$ fusion rule.  Fig.~\ref{fig:EvolutionN} depicts the evolution of the problem corresponding to Alexis's decision $\Hhat_1$.

    \begin{figure}[t]
        \centering
        \includegraphics[width=3.4in]{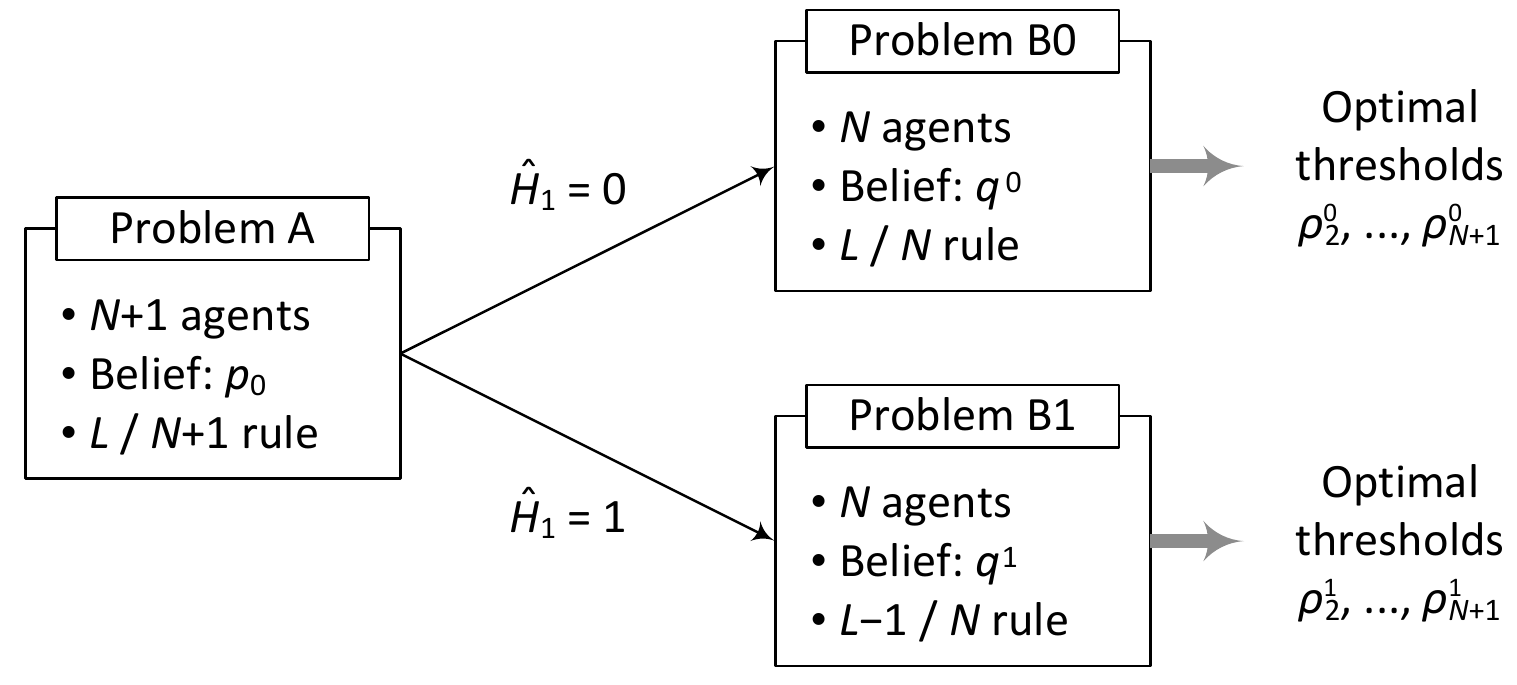}
        \caption{An $(N{+}1)$-agent problem is divided into two $N$-agent problems depending on Alexis's decision $\Hhat_1$.}
        \label{fig:EvolutionN}
    \end{figure}

    Let us find the optimal thresholds $\rho_2^{_0 \ast}, \ldots, \rho_{N+1}^{_0 \ast}$ in Problem B0 in Fig.~\ref{fig:EvolutionN}.  In fact, the $N$ agents in Problem B0 also observe public signals.  However, because of the condition that the existence of the public signals does not affect optimal decision thresholds of a team of $N$ agents for any $K$-out-of-$N$ fusion rule, we can find the optimal thresholds as if the agents do not observe the public signals.  Since a decision of an agent is critical only if the other $N{-}1$ local decisions are $L{-}1$ ones and $N{-}L$ zeros, the optimal decision threshold $\rho^{_0 \ast}$ is the solution to
    \begin{align}
        \frac{f_{Y | H} (\rho^{_0} \MID 1)}{f_{Y | H} (\rho^{_0} \MID 0)}
        & = \frac{c_{10} \Belief^{_0} \binom{N-1}{L-1} \left(\LocalIxSeq{0} \right)^{L-1} \left(1{-}\LocalIxSeq{0} \right)^{N-L}}{c_{01} (1{-}\Belief^{_0}) \binom{N-1}{N-L} \left(\LocalIIxSeq{0} \right)^{N-L} \left(1{-}\LocalIIxSeq{0} \right)^{L-1}} \nonumber \\
        & = \frac{c_{10} p_0 \left(1{-}\LocalIx \right) \left(\LocalIxSeq{0} \right)^{L-1} \left(1{-}\LocalIxSeq{0} \right)^{N-L}}{c_{01} p_1 \LocalIIx \left(\LocalIIxSeq{0} \right)^{N-L} \left(1{-}\LocalIIxSeq{0} \right)^{L-1}},
        \label{eq:DecThres0,Nagents,serial}
    \end{align}
    where $\Belief^{_0}$ is replaced by \eqref{eq:BeliefUpdate_N+1}.  Due to the assumption that $\rho_1^{\ast} = \lambda_1^{\ast}$, $\LocalIx$ and $\LocalIIx$ in \eqref{eq:DecThres0,Nagents,serial} are the same as $\LocalIx$ and $\LocalIIx$ in \eqref{eq:DecThres,N+1agents,parallel}.

    Comparing \eqref{eq:DecThres0,Nagents,serial} to \eqref{eq:DecThres,N+1agents,parallel}, we can find that they have the same solutions, i.e., $\rho_i^{_0 \ast} = \lambda_i^{\ast}$.  Therefore, the agents should not change their decision thresholds after observing $\Hhat_1 = 0$.

    We can also find the optimal thresholds $\rho_2^{_1 \ast}, \ldots, \rho_{N+1}^{_1 \ast}$ in Problem B1 in Fig.~\ref{fig:EvolutionN} by looking at the $N$-agent problem without public signals:
    \begin{align}
        \frac{f_{Y | H} (\rho^{_1} \MID 1)}{f_{Y | H} (\rho^{_1} \MID 0)}
        & = \frac{c_{10} \Belief^{_1} \binom{N-1}{L-2} \left(\LocalIxSeq{1} \right)^{L-2} \left(1{-}\LocalIxSeq{1} \right)^{N-L+1}}{c_{01} (1{-}\Belief^{_1}) \binom{N-1}{N-L+1} \left(\LocalIIxSeq{1} \right)^{N-L+1} \left(1{-}\LocalIIxSeq{1} \right)^{L-2}} \nonumber \\
        & = \frac{c_{10} p_0 \LocalIx \left(\LocalIxSeq{1} \right)^{L-2} \left(1{-}\LocalIxSeq{1} \right)^{N-L+1}}{c_{01} p_1 \left(1{-}\LocalIIx\right) \left(\LocalIIxSeq{1} \right)^{N{-}L{+}1} \left(1{-}\LocalIIxSeq{1} \right)^{L{-}2}}.
        \label{eq:DecThres1,Nagents,serial}
    \end{align}
    Again, due to the assumption that $\rho_1^{\ast} = \lambda_1^{\ast}$, $\LocalIx$ and $\LocalIIx$ in \eqref{eq:DecThres1,Nagents,serial} are the same as $\LocalIx$ and $\LocalIIx$ in \eqref{eq:DecThres,N+1agents,parallel}.  We reach to the same statement that the two equations have the same solutions, i.e., $\rho_i^{_1 \ast} = \lambda_i^{\ast}$, by comparing \eqref{eq:DecThres1,Nagents,serial} to \eqref{eq:DecThres,N+1agents,parallel}.  Thus, the agents should not change their decision thresholds after observing $\Hhat_1 = 1$.

    Consequently, for a team of $N+1$ agents and any $L$-out-of-$(N+1)$ rule, their optimal decision thresholds are the same whether they observe previous decisions or not.
\end{IEEEproof}

We want to clarify the term \emph{belief} in Fig.~\ref{fig:EvolutionN}, which is distinguished from prior probability.  The prior probability is fixed as $p_0$ and known to all agents.  The belief is how probable the agents think $H = 0$ is.  The belief can change as agents observe previous decisions even though the prior probability cannot.  The above proof does not mean that the prior probability is changed after the agents observe Alexis's decision.  It is saying that the problem of determining optimal thresholds upon observing Alexis's decision is equivalent to Problem B0 or B1 corresponding to $\Hhat_1$.

The essence of this proof is to show that the effect of new fusion rules, $L$-out-of-$N$ and $(L{-}1)$-out-of-$N$, exactly cancels out the effect of new beliefs $\Belief^{_0}$ and $\Belief^{_1}$.  It is observed when we compare \eqref{eq:DecThres0,Nagents,serial} and \eqref{eq:DecThres1,Nagents,serial} to \eqref{eq:DecThres,N+1agents,parallel}.  As we mentioned earlier in Section~\ref{sec:Introduction}, the public signals seem to have the same amount of a positive feedback and a negative feedback.  From an overall standpoint, the public signals neither harm nor help the decision-making task.

\begin{corollary}
    Suppose that sharing previous decisions does not change Alexis's decision rule (i.e., $\rho_1^{\ast} = \lambda_1^{\ast}$).  For any $N$ and $L$-out-of-$N$ fusion rule, the existence of the public signals does not affect optimal decision thresholds of a team of $N$ agents.
    \label{cor:GeneralN}
\end{corollary}

\begin{IEEEproof}
    By mathematical induction, it is true from Theorems~\ref{thm:2Agent} and~\ref{thm:NtoN+1}.
\end{IEEEproof}

Corollary~\ref{cor:GeneralN} is trivial for $N = 1$ and is proven by Theorem~\ref{thm:2Agent} for $N = 2$.  For $N \geq 3$, we can iteratively break down the problem until we have $2^{N - 2}$ two-agent problems like in Fig.~\ref{fig:EvolutionN}.  The backward process from the leaves (two-agent problems) to the root ($N$-agent problem) will demonstrate Corollary~\ref{cor:GeneralN}.

Corollary~\ref{cor:GeneralN} only refers to the optimal decision thresholds.  However, it implies that team's performance (i.e. Bayes risk) is not affected by the public signals.  This is because decision thresholds determine the probabilities of errors and, thus, the Bayes risk.

\section{Conclusion}    \label{sec:Conclusion}

We have discussed sequential decision making in a distributed detection and fusion system by a team of agents.  It would be intuitively desired to have as much information as possible when performing a hypothesis test.  However, our study has revealed that it is useless to observe other agents' decisions in this system.  The agents just need to individually make the best decisions.

This result is justified by symmetries throughout our model.  The first symmetry is the equal quality of private signals due to the identical likelihood functions.  The second symmetry is the equal (1-bit) votes that agents have.  The third symmetry is the equal weights of the votes implied by the $L$-out-of-$N$ fusion rule.

One interpretation of the main result is as follows:
Each agent sends 1 bit of information about its private signal to the fusion center.  If Blake refers to Alexis's decision, then the fusion center effectively receives less than 1 bit of information about Blake's private signal from him.  Therefore, in order to prevent such an efficiency loss, Blake should make a reasonable decision only based on his own private signal.

This result is obtained under an assumption that Alexis uses the same decision thresholds in both scenarios, and the assumption is not unreasonable.  The assumption is trivially true for $N = 1$.  It is also true for $N = 2$;  our proof of Theorem~\ref{thm:2Agent} even does not use this assumption.  In addition, we have confirmed that it is true at least for any $N \leq 9$ by numerical experiments.  This assumption heuristically seems true, and our future work is to verify it for arbitrary $N$.

\section*{Acknowledgment}
Discussions with V. Krishnamurthy, J. Z. Sun, and L. R. Varshney are greatly appreciated.

\bibliographystyle{IEEEtran}
\bibliography{rhim_lib}
\end{document}